# Outdoor 100-m-scale air waveguides

**Author List**

Andrew Goffin[1,2], Gregory Babic[1], Andrew DeRusha[2], Maksim Livshits[2,*], Howard M. Milchberg[1,*], Patrick Skrodzki[2,*]

**Affiliations**

[1] Institute for Research in Electronics and Applied Physics, University of Maryland, College Park, Maryland 20742, USA

[2] Physical Chemistry and Applied Spectroscopy, Los Alamos National Laboratory, Los Alamos, NM 87544, USA

*Corresponding authors: mlivshits@lanl.gov, milch@umd.edu, pskrodzk@lanl.gov

**Abstract**

**Optical power densities for standoff spectroscopy, remote sensing, directed energy, and free-space optical communications are limited by diffraction and adverse atmospheric conditions such as turbulence, fog, and wind. Air waveguides, generated by ultrashort-pulsed laser filamentation, are a promising approach for transmission and collection of optical signals over long distances, overcoming beam diffraction and, for point-like sources, inverse square signal falloff with distance. In this work, we demonstrate outdoor guiding for the first time over a record length of ~100 m, exceeding the prior (indoor) record of 42 m. We correlate air waveguide performance to a range of real atmospheric conditions including turbulent refractive index structure parameter ($C_n^2$) values up to ~2.5×10$^{-14}$ m$^{-2/3}$, crosswind speeds up to ~2 m/s, as well as temperature, pressure, and humidity variations. Accompanying propagation simulations provide insight into the real effects of these environmental perturbations, particularly crosswind, on waveguide performance and lifetime. Our results pave the way for quasi-continuous air waveguiding with kHz-scale repetition rate filaments in challenging outdoor environments.**

Air waveguides (AWGs) formed by femtosecond optical filaments are a recently developed technology[1–3] with potential applications to remote sensing[4], directed energy[5], and free space optical communications[6]. Air waveguides are formed by inducing an annular air density depression that follows in the wake of filamentation using a spatially structured femtosecond laser pulse heating the air. In the original demonstration[1] of a $\sim 1\ \mathrm{m}$ waveguide, the structure was produced by filamentation of a four-lobed $\mathrm{TEM}_{11}$-like beam that generated an array of four femtosecond filaments. Energy deposited by the filaments heated the air, leaving reduced air density (and therefore refractive index) regions. After thermal diffusion, the individual density depressions merged into a low-density annulus surrounding a relatively unperturbed central region, forming the waveguide cladding and core, with the resulting index profile analogous to that of a graded-index optical fiber. Subsequent work extended this method to 50-m-long waveguides generated by high-energy $\mathrm{LG}_{01}$ "donut" beams; the donut beam generates a dense pattern of multiple filaments around the intensity ring, producing a more continuous annular heated region from the overlap of dense filament cores (described below), which is better suited for long-distance guiding[2]. Air waveguide lengths are limited by the distance over which the structured beam maintains sufficient intensity, so that an adequate number and density of filament cores survive around the annulus to form a uniform waveguide cladding. More recently, high-repetition-rate patterned filamentation over $\sim$1 m has been used to sustain the thermally generated cladding against diffusion, enabling quasi-steady-state air waveguiding[3].

Femtosecond laser filamentation manifests from the interplay between diffraction and competing nonlinear self-focusing and defocusing effects. In air, self-focusing arises mainly from the nonlinear polarization of $N_2$ and $O_2$ molecules, which have a near-instantaneous electronic and delayed rotational response to the strong field of an intense, ultrashort laser pulse. The result is a pulse width-dependent nonlinear refractive index $n_{2,eff}(\tau)$.[7,8] Self-focusing collapse occurs for laser peak power $P > P_{cr} = 3.77\lambda^2/8\pi n_{2,eff} n_0$[9,10], where $P_{cr}$ is pulse width dependent and ranges over $\sim 2.5 - 12$ GW at $\lambda = 800$ nm, until arrested by low-density ($N_e \sim 10^{16}$ cm$^{-3}$) plasma generation. The dynamic competition among self-focusing, plasma defocusing, and diffraction gives rise to the optical filament structure, with the intensity clamped to $I \sim 10^{14}$ W/cm$^2$ within a "core" of diameter $d_{\mathrm{core}} \sim$ 200 µm. Filament propagation distances are much longer than the Rayleigh range associated with the core ($L_{fil} \gg \pi d_{core}^2/4\lambda$). Surrounding the core is a weaker electromagnetic "reservoir" that replenishes the core[11] via spatiotemporal optical vortex-mediated energy flow[12,13]. The filament length roughly scales with the Rayleigh range of its reservoir, so larger beams will generate longer filaments[14]. Filaments appear robust against aerosol blockages[15] and turbulence[16], making them suitable candidates for propagation through adverse atmospheric conditions. For pulses with $P \gg P_{cr}$, amplified noise in the beam intensity profile can nucleate into multiple filament cores[17–19].

Filamenting pulses deposit energy through plasma generation[20], molecular rotational excitation[21–23], and electron-neutral collisions[24]. Plasma recombination and thermalization occurs on $\sim 10$ ns timescales, much quicker than the air's fastest acoustic response time $d_{core}/2c_s \sim 300$ ns, where $c_s \sim 300$ m/$s$ is the air sound speed. This creates a filament-centered hot ($\Delta T \sim 100$ K) pressure spike, which equilibrates by radially launching a single-cycle acoustic wave, leaving behind a gas density depression or "density hole"[25–27]. The density hole then slowly dissipates over tens of milliseconds via thermal diffusion in air. In prior air waveguide experiments[1–4], it is an annular array of such density holes, merged via thermal diffusion, that forms the waveguide cladding, with the central less-perturbed air serving as the waveguide core. We note that previous work has simulated[28,29] and measured[1,30] confined propagation of microwave pulses in a ring of filament plasmas and optical pulses in filament-generated acoustic waves. However, those guides have extremely short lifetimes limited by plasma recombination[30] ($\sim$10 ns) and acoustic[1] ($\sim$0.1 μs) timescales and, in the case of the filament plasma guide, limited capability to confine optical radiation due to large gaps between the filament plasmas. Figure 1(a) compares the single-shot lifetimes and estimated core-cladding index contrasts $\Delta n_{\mathrm{vis}}$ of these air-based guiding structures assuming six filaments in a ring, air sound speed $c_s = 343$ m/s, and thermal diffusivity $\alpha = 0.217$ cm$^2$/s.

In this paper, we demonstrate air waveguiding in outdoor conditions for the first time, successfully transitioning the method from indoor propagation ranges[1–3] with a minimal loss of fidelity. We extend these outdoor waveguides to record lengths of $\sim$100 m, proving that the distance scalability demonstrated in prior indoor work[2] also applies to outdoor settings, and paving the way for a fieldable platform for remote optical sensing, directed energy, and free-space optical telecommunications.

The outdoor range and experiment setup are located at Los Alamos National Laboratory (see Figure 1(b)(i-iii)). Filamentation and air waveguides are driven by a Ti:sapphire-based system inside a container unit at the field site, with 210 mJ after the final optic of the telescope, central wavelength $\lambda = 800$ nm, 10 Hz repetition rate, and the pulse stretched to $\tau = 1.6$ ps (FWHM). The pulse duration maximizes the filament propagation range by increasing the number of refocusing cycles induced by the molecular rotational response[8]. The laser output beam is formed into a near-$LG_{01}$ mode using an $\ell = +1$ spiral phase plate and down-collimated to a ring diameter $d_{ring} = 8\,\mathrm{mm}$, which sets the initial size of the waveguide and linear Rayleigh range $z_R = \pi d_{ring}^2/2\lambda \sim 130$ m. The donut beam is launched from the container unit, with multiple filaments forming around the circumference[2] by the container unit exit ($z \sim 0$ m). We inject a 5-mW, green (520 nm) continuous-wave probe laser diode through the telescope at f/1200 ($z_R \sim 1$ m) into the air waveguide to measure guided signal enhancement and guiding timescales.

Diagnostics for this experiment include (1) burn paper to capture the arrangement of filaments which generate the waveguide cladding along the beam path, (2) an off-axis camera and beam block translated along the propagation path to image the guided and unguided probe beam, (3) a photodiode to measure the time-dependent probe beam transmission, and (4) a scintillometer and hot-wire anemometer to measure atmospheric turbulence (air structure constant $C_n^2$), humidity, pressure, and crosswind speed. Representative data collected from diagnostics (2) and (3) are shown in Fig. 1(b)(iv,v) and diagnostic details are covered in Methods.

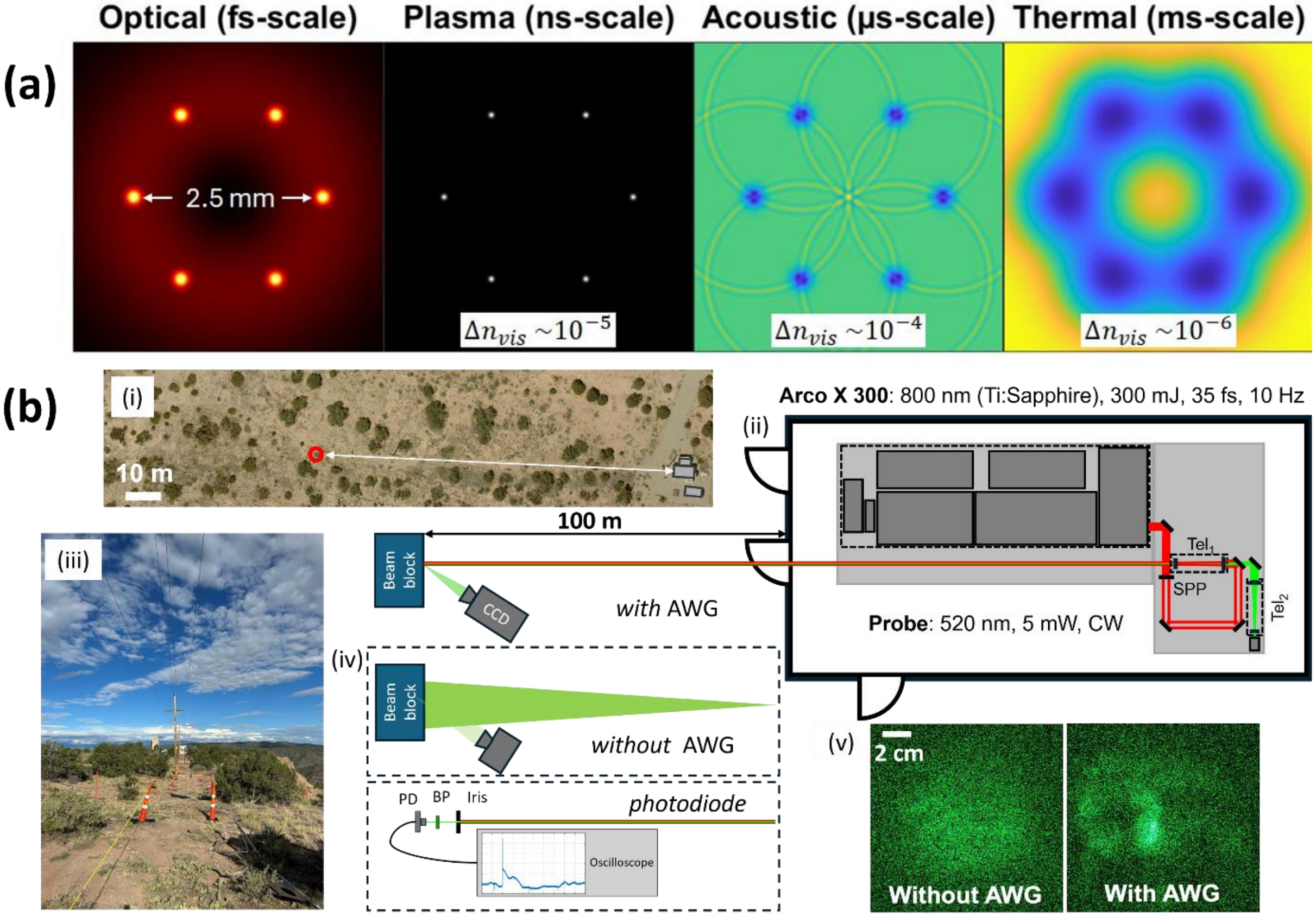

**Fig. 1. (a)** Illustration of different timescales in air waveguide formation from filaments, from shortest (optical pulse, fs-scale) to longest (thermal waveguide, ms-scale), with a sample $d_{ring} = 2.5$ mm $LG_{01}$ beam. The optical image shows a multi-filamenting $LG_{01}$ beam fluence profile, the plasma image shows the electron density profile from a ring of filaments, and the acoustic and thermal images show the air density profile in each case. $\Delta n_{vis}$ is the approximate core-cladding contrast for visible light in each waveguide. **(b)** Illustration of setup at field site for experiment. *(i):* Overhead view of the 100m range. *(ii):* Experimental diagram with laser beamline. SPP: Spiral Phase Plate, $Tel_1$: Down-collimation telescope for 800nm filamenting beam, $Tel_2$: Telescope for 520nm coupling into the waveguide. *(iii)*: View back from $z = 100$ m to laser source ($z = 0$). *(iv):* Diagrams showing beam spreading without the waveguide ("No AWG") and the experimental setup for photodiode measurements ("Photodiode"). *(v)*: Images of green beam with and without the waveguide at $z = 100$ m.

We conduct experiments over several days (August 27, August 28, September 25, September 30, and October 1, 2025) and different times of day, including night, to ensure reproducibility and measure performance over a wide range of environmental conditions. In all cases, we keep the operational parameters of the 800-nm waveguide laser and 520-nm probe laser constant, so that the main variable is day-to-day atmospheric changes.

**Results**

The performance of the air waveguide along the propagation path is assessed in Fig. 2. The burn patterns in Fig. 2(a) show the arrangement of filament cores, which comprise the air waveguide cladding, from a single driving laser shot at several distances. Most of the filament cores do not survive at distances beyond 80 m; however, as shown by Fig. 2(b), the waveguide generated by the few surviving filaments is sufficient to sustain guiding to 100 m. Counting the number of filament cores at the distance of 10 m is difficult, since the reservoir damages the burn paper here; nevertheless, we estimate the number of filaments using previous work[2]. For pulse duration $\tau = 1.6\ \mathrm{ps}$ and field site ambient pressure ~800 hPa, the nonlinear index is[7,8] $n_{2,eff} \approx 3.2 \times 10^{-19}$ W/cm$^2$. This gives an estimated[2,31] $n_{fil} \approx \sqrt{2.9\varepsilon_{LG}/\varepsilon_{cr} - 3} \sim 11$ filaments for an $LG_{01}$ mode of energy $\varepsilon_{LG} \sim 210$ mJ and $\varepsilon_{cr} = P_{cr}\tau$, where $P_{cr} \sim 3$ GW. After initial thermal diffusion of the density holes to form the cladding, the effective air waveguide core diameter is ~10 mm. By 100 m, we observe that 3 to 4 filaments typically survive, where the maximum diameter across the structure is ~12 mm.

Figure 2(b) compares the green probe beam images with the waveguide off and on, showing a representative shot at each distance. For each image the camera uses a 0.5-ms exposure while synchronized to the AWG 10 Hz laser system. The unguided probe beam diameter is larger than the waveguide core diameter at distances greater than ~10 m. In the waveguide-on images, we observe guided light confined to the core as well as several circular shadows caused by probe refraction from individual density holes. The probe is strongly defocusing, such that it becomes dimmer with distance without the presence of the waveguide to confine energy to the core. At some distances, the images are collected during the day in bright sunlight; as such, the waveguide-off images appear particularly noisy at longer distances (80 m and 100 m). Even so, guiding is apparent, as evidenced by a bright spot whose size corresponds to ~10 mm diameter at 100 m.

Figure 2(c) plots the air waveguide signal enhancement with distance. Signal enhancement is defined as $\epsilon = (\bar{E}_g - \bar{E}_{Ng})/(\bar{E}_{ug} - \bar{E}_{Nug})$, where $\bar{E}_g$ is the mean probe beam energy within $D_{core} \sim 10\ \mathrm{mm}$ for the waveguide on (guided), $\bar{E}_{ug}$ is mean probe energy in the same area for the waveguide off (unguided), and $\bar{E}_{Ng}$ and $\bar{E}_{Nug}$ are the mean noise/background signals for no probe and waveguide present, collected immediately after $\bar{E}_g$ and $\bar{E}_{ug}$ respectively. In practice, we found that $\bar{E}_{Ng} \approx \bar{E}_{Nug}$ due to the noise being dominated by sunlight which was similar for both datasets in all cases. At each distance, 100 images are recorded to determine each of the mean value, with error bars corresponding to $\sigma_\epsilon = \pm\epsilon\sqrt{\sigma_{E_g}^2/\bar{E}_g^2 + \sigma_{E_{ug}}^2/\bar{E}_{ug}^2}$ for the $E_g$ variance, $\sigma_{E_g}^2$, and $E_{ug}$ variance, $\sigma_{E_{ug}}^2$. It is seen that $\epsilon$ increases with distance to ~1.5 at $100\ \mathrm{m}$. To assess the signal detectability, we calculate the increase in the signal-to-noise ratio, $\Delta\mathrm{SNR} = \log_{10}[(\bar{E}_g - \bar{E}_{Ng})/\sigma_{Ng}] - \log_{10}[(\bar{E}_{ug} - \bar{E}_{Nug})/\sigma_{Nug}]$, where $\sigma_{Ng}$ and $\sigma_{Nug}$ are standard deviations of the noise signal. Here, the plotted error bars are ± the standard deviation of ΔSNR, computed from the standard deviation of the first and second terms of ΔSNR added in quadrature. ΔSNR increases from ~0.8 dB to ~2 dB over the propagation range, with significantly increasing error bars at the longest distances stemming from the accumulation of atmospheric-induced effects on detection.

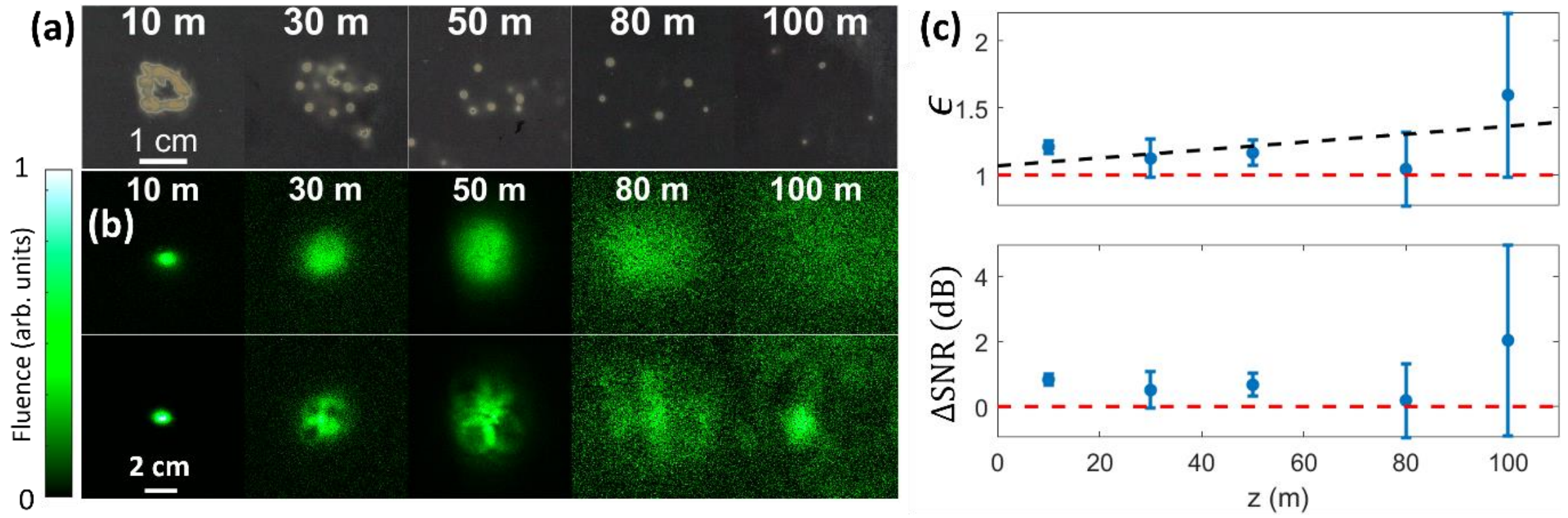

**Fig. 2.** Demonstration of guiding over 100m outdoors. **(a)** Burn paper images showing filamentation at five different locations along the beam path. **(b)** Single-shot images of the guided beam at five locations along beam path (delay = 1 ms). The color bar is set so 1 is the maximum pixel value for each distance. **(c)** *Top:* Signal enhancement $\epsilon$ in the core region as a function of distance. The black dashed line corresponds to a linear fit of the mean $\epsilon$ vs. $z$. *Bottom*: SNR gain ΔSNR as a function of distance. Both plots average over 100 shots at 1 ms delay.

Next, we assess the time-dependent performance of the air waveguide at the longest guiding distance, $z = 100$ m. Figure 3(a) shows the guided green beam at increasing delay; each image corresponds to a 0.5-ms camera exposure synchronized to the waveguide-generating laser to follow filamentation. Time-dependent signal enhancement $\epsilon(t)$ and ΔSNR, plotted in Fig. 3(b), are determined from the images as in Fig. 2(c). The useful lifetime of the waveguide is seen to be < 3 ms, much shorter than the ~100 ms estimated for this waveguide size based on prior results[2]. To verify this timescale, we used a photodiode on the probe optical axis to measure the time-dependent signal in the core region with(out) the waveguide $\mathcal{E}_g(t)(\mathcal{E}_{ug}(t))$ with background signal $\overline{\mathcal{E}}_{Ng}(\overline{\mathcal{E}}_{Nug})$ and calculate signal enhancement, defined as $\epsilon_{PD}(t) = (\overline{\mathcal{E}}_g(t) - \overline{\mathcal{E}}_{Ng})/(\overline{\mathcal{E}}_{ug}(t) - \overline{\mathcal{E}}_{Nug})$. The average signal enhancement is plotted in Fig. 3(c) as the black line, where all the quantities in the expression for $\epsilon_{PD}(t)$ are averages of 50 photodiode traces. Each 0.25-ms interval is subjected to a two-sample t-test to determine the likelihood of guiding ($\epsilon_{PD} > 1$). The data passes the t-test until $2.75 - 3$ ms, where the lower confidence bound on $\epsilon_{PD}$ crosses below unity (shown by the orange region in Fig. 3(c)).

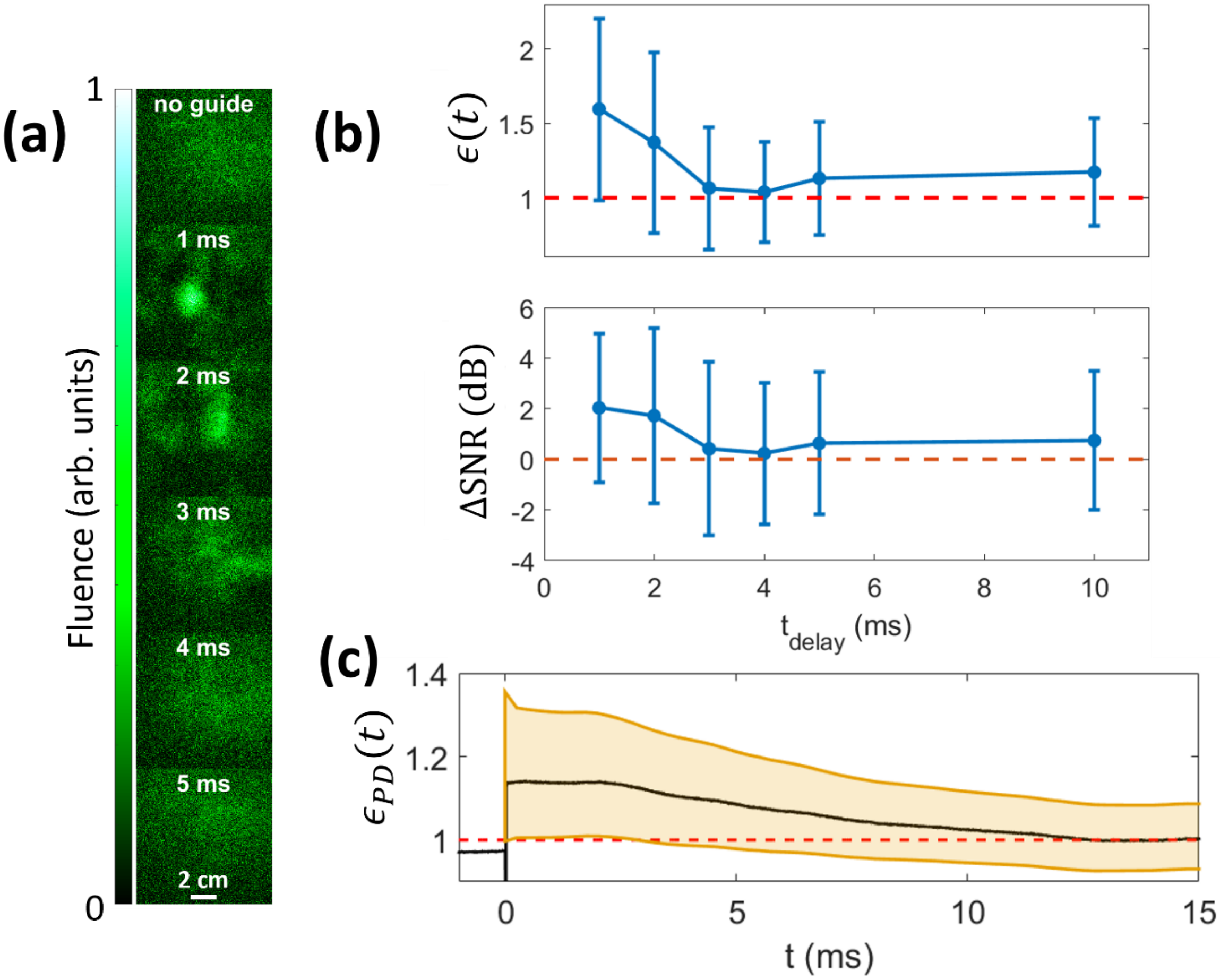

**Fig. 3. (a)** Guide images at $z = 100$ m versus delay with a 0.5 ms exposure time on September 25. **(b)** Calculated waveguide signal enhancement $\epsilon(t)$ (*top row*) and SNR increase $\Delta$SNR (*bottom row*) from guide images. **(c)** *Black trace:* Average photodiode trace of 50 shots collected on October 1, $v_{wind} \approx 1.6$ m/s. *Orange shaded region*: Confidence interval for $p < 0.05$ in the two-sample t-test in 0.25 ms increments. *Red dashed line:* Marks $E_g = E_{ug}$, where the waveguide is non-functional.

Finally, we assess waveguide survivability in real atmospheric conditions. In Fig. 4, we report four different data sets taken at $z = 100$ m during different days and times of day. Figure 4(a) shows sample time-varying environmental data collected on September 30. On this date, there were large wind gusts between 15:30 and 16:30 local time, leading to high values of wind speed $v_{wind}$ and refractive-index structure parameter $C_n^2$, a measure of turbulence strength. Guided modes for three dates, August 27, August 28, and September 25, are shown in Fig. 4(b). The data on August 27 was collected at night, leading to lower sun-dominated noise than on the other dates. Additionally, the quality of the waveguide on August 27 was relatively poor, as shown by the stronger density holes on the top right and bottom left compared to the other corners. Waveguide performance for each date is shown in Fig. 4(c), alongside the crosswind speed $v_{wind}$, with signal enhancement $\epsilon$ plotted versus delay. A notable feature includes the rapid drop in $\epsilon$ at $\sim$3 ms on September 25. In contrast, on August 27 the signal enhancement remains $\epsilon > 1$ for $t \leq 5$ ms, consistent with the slowest wind on that date. August 28 has later delays omitted due to the shifting alignment, although early-time $\epsilon$ is slightly higher than it is on other dates and with lower standard deviation than September 25.

The shot-to-shot pointing variation for the three 100-image datasets (color-coded as in Fig. 4(c)) is shown in Fig. 4(d) by plotting the guided probe beam centroid $(x, y)$ of every image at 1-ms delay (top) and their corresponding standard deviations $(\sigma_x, \sigma_y)$ (bottom). Here, the crosses and solid circles correspond to unguided and guided beams, respectively. Interestingly, the August 28 and September 25 data show that the centroid variation is higher with the guide than without it, likely caused by drifting of the laser-imprinted density depression in the wind direction. Without the waveguide, the beam is weakly deflected by turbulence ($C_n^2 < 10^{-14}$ m$^{-2/3}$). Table I shows the average environmental parameters for each day during data collection.

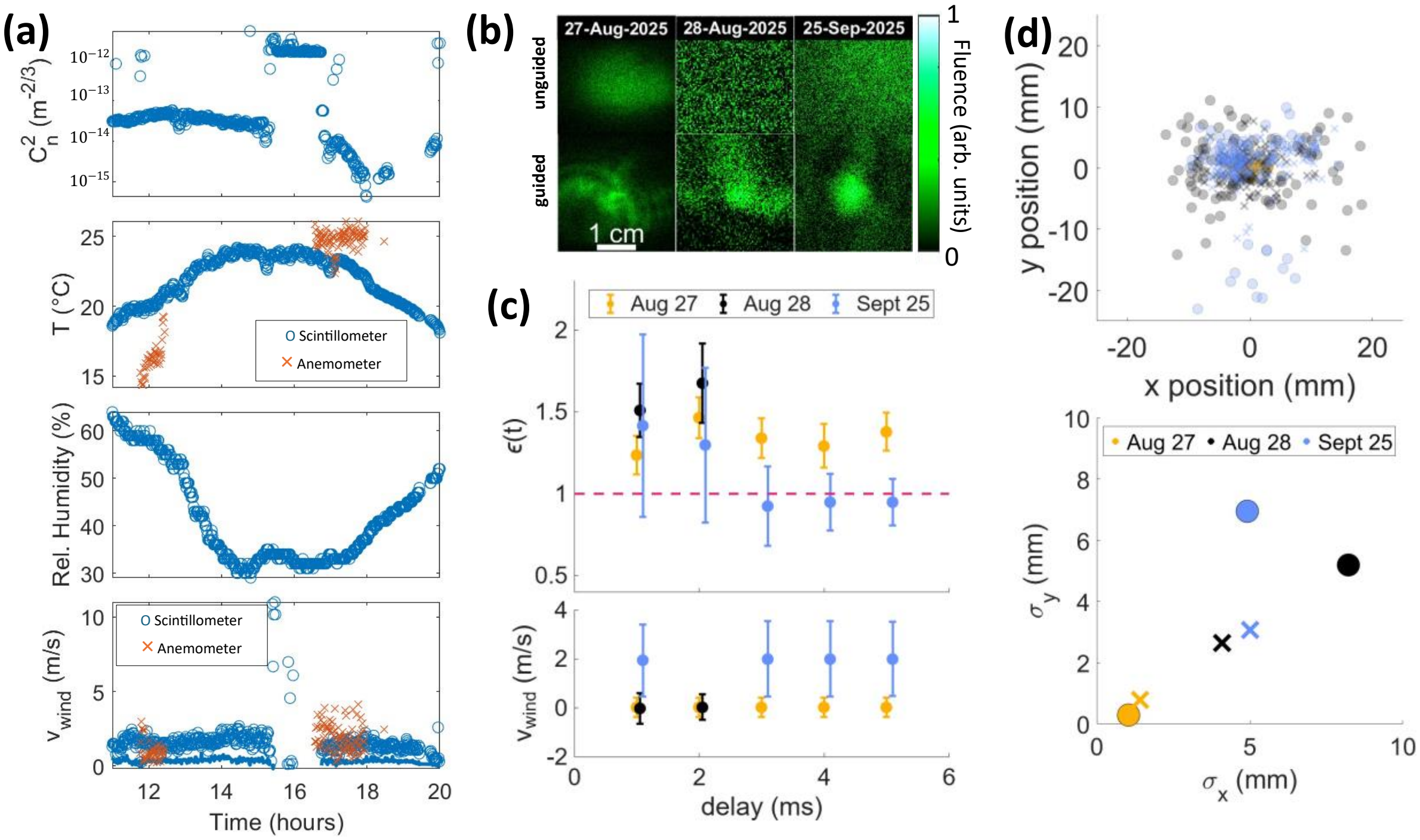

**Fig. 4.** Guiding with different environmental parameters. **(a)** Sample environmental data from September 30. **(b)** Sample mode images from three days, top row being the unguided probe and bottom row being guided. **(c)** *Top:* Signal enhancement $\epsilon(t)$ vs delay for each day. *Bottom:* Measured wind speed $v_{wind}$ (with error bars corresponding to standard deviation) for each day and delay, based on scintillometer and anemometer data. **(d)** *Top:* Shot-to-shot beam centroid for each date without the guide (circle) and with the guide (×). *Bottom*: Average beam centroid variance for each date without the guide (circle) and with the guide (×). Each plot uses 1 ms delay data.

| Date | $C_n^2$ (m$^{-2/3}$) | Temp (˚C) | Rel. humidity (%) | $\langle v_{wind} \rangle$ (m/s) | $\sigma_{v_{wind}}$ (m/s) |
|---|---|---|---|---|---|
| Aug. 27 | 0.11×10$^{-14}$ | 19.3 | 69 | 0.03 | 0.4 |
| Aug. 28 | 2.43×10$^{-14}$ | 29.1 | 30 | 0.07 | 0.83 |
| Sept. 25 | N/A | 28.3 | N/A | 2.00 | 1.52 |

**Table I.** Reported environmental parameters for each dataset represented in Fig. 4(b-d). On Aug. 27 and Aug 28, the data was recorded with a scintillometer operating over the 100 m range. On Sept. 25, the scintillometer was not operational and a hot-wire anemometer was used to report crosswind speed and air temperature.

**Discussion**

A key difference in these experiments from prior indoor efforts[2] is the presence of wind. In Fig. 5, we analyze the impact of wind on air waveguide performance by modeling waveguide displacement and the subsequent guided mode. We assume that the waveguide and probe are aligned at 1-ms delay, matching our experimental alignment procedure. Fig. 5(a) shows the density hole profiles versus propagation distance for varying right-directed wind velocities. The density hole distribution versus $z$ is imposed to approximately match burn patterns (Fig. 2(a)) and to start with the number of filaments $n_{fil} = 11$ for the beam power estimated earlier. All density holes in Fig. 5(a) are shown with 5-ms delay, so they move to the right at 5 mm per 1-m/s wind speed; we assume that the full length of the waveguide moves sideways rigidly. Figure 5(b) shows the simulated guided modes (see Methods) at $z = 100\ \mathrm{m}$ for three delays. At 2-ms delay, the probe beam is guided for all wind speeds, but at the longer delays, higher wind speeds increasingly push the waveguide out of the way of the probe and reduce guiding.

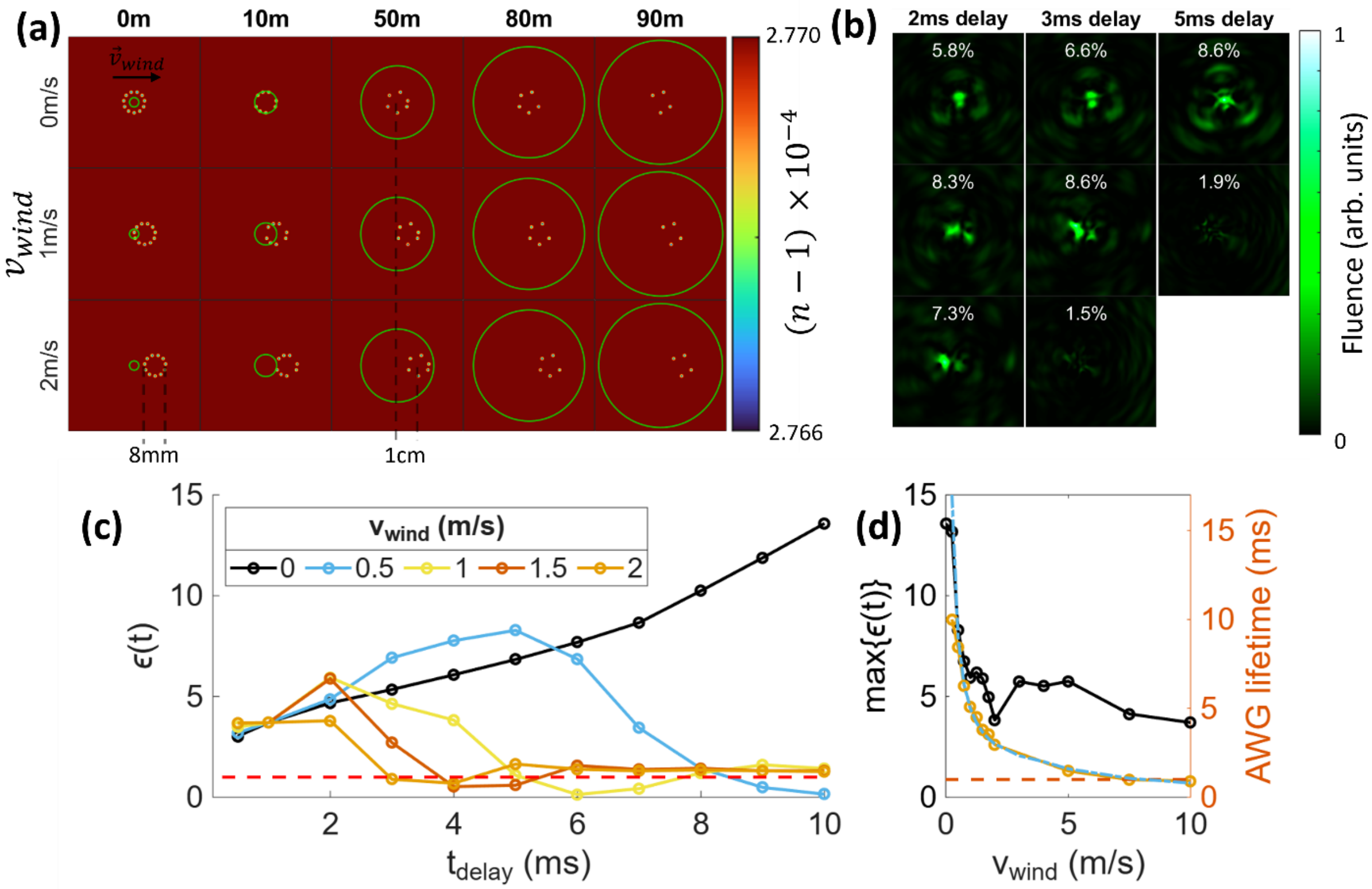

**Fig. 5. (a)** Density hole profiles at 5 ms delay for different propagation distances (columns) and wind speeds $v_{wind}$ (rows). The green circles enclose 50% of the unguided beam energy at each position. **(b)** Simulated guided beam profiles at 100m for different delays (columns) and wind speeds (rows). The percentage in each image is the percentage of the total beam energy in the waveguide core (compared with ~1.9% without the waveguide). **(c)** Simulated waveguide signal enhancement $\epsilon$ for different wind speeds vs. delay. The red dashed line marks $\epsilon = 1$. **(d)** Simulated maximum signal enhancement $\max\{\epsilon(t)\}$ and waveguide lifetime vs. wind speed. The orange dashed line corresponds to a waveguide lifetime of 1 ms, and the lifetime curve is overlaid with the blue dash-dot lifetime estimate of $\Delta t_{life} = (d_{ring}/2 + \min\{x_0, d_{ring}/2\})/v_{wind}$.

Figure 5(c) plots the signal enhancement $\epsilon(t)$ vs. delay of several simulated wind speeds. Without wind, the waveguide lifetime (the duration over which $\epsilon > 1$) would be limited by thermal diffusion[2] to ~100 ms. However, wind moves the waveguide off the optical axis through advection, decreasing the lifetime, as clearly seen in Fig. 5(c), where $\epsilon$ crosses unity earlier as wind speed increases. In Fig. 5(d), the maximum $\epsilon(t)$ is plotted vs. wind speed (left scale), which rapidly decreases for $v_{wind} \rightarrow \sim 2$ m/s and remains roughly constant for $v_{wind} > 2$ m/s. The air waveguide (AWG) lifetime from simulations (right scale) is plotted as orange circles. We also consider a simple lifetime model, $\Delta t_{life} = (d_{ring}/2 + \min\{x_0, d_{ring}/2\})/v_{wind}$, where $\Delta t_{life}$ is the waveguide lifetime and $x_0 = v_{wind} t_{delay}$ is the starting waveguide center (with centered delay $t_{delay} = 1$ ms). This equation gives the blue dashed curve in Fig. 5(d), which closely matches the simulated lifetime. The simulations also predict possible further optimization: the curve in Fig. 5(c) predicts that peak enhancement is $\epsilon = 3.8$ at $v_{wind} = 2$ m/s, suggesting that the experimental setup can be further optimized beyond the measured $\epsilon = 1.5$ at $v_{wind} = 2$ m/s.

The blue curve in Fig. 5(d) also suggests that quasi-steady state air waveguides generated by high rep. rate filamentation, which have been previously demonstrated to beat the thermal dissipation lifetime limit[3], can also mitigate the effects of wind. The curve predicts $\Delta t_{life}$~3 ms for experiments with $v_{wind} = 2$ m/s, which matches well the measured lifetime of 2–4 ms for the same $v_{wind}$. Using this trend, for wind speeds below 7.5 m/s (17 mph) we predict a quasi-steady-state guiding regime with a 1-kHz repetition rate filament-driver laser replenishing the waveguide faster the wind drift speed.

From the results of our experiments and simulations, we conclude that air waveguide performance mainly depends on (a) adequate azimuthal coverage of the filament ring (and hence waveguide cladding) and (b) crosswind speed. In the Aug. 27 data in Fig. 4, although the overall efficiency is lower due to poor filament coverage around the ring, the lifetime of the waveguide is notably longer, with $\epsilon = 1$ reached only after 5-ms delay. This is attributed to the lower wind speed of $v_{wind} < 0.4$ m/s (Fig. 4(c)), which is associated with $\Delta t_{life} \geq 10$ ms (Fig. 5(d)). Other environmental factors, such as temperature, humidity, and turbulence, can play a role. However, the waveguides on August 28 and September 25 had noticeably different performance at early delays despite similar air temperatures. Additionally, in-lab experiments[32] indicate that air turbulence at typical outdoor values is not detrimental until kilometer ranges due to scintillation of the filamenting ring. In our case, the most asymmetric filament ring profile was generated in the *weakest* turbulence dataset, which indicates a systematic error as opposed to turbulence-induced scintillation. Additionally, supplemental simulation modelling of guiding through turbulence at $C_n^2 = 10^{-13}$ m$^{-2/3}$ (higher than Table I values by $> 4\times$) shows average $\epsilon$ variance of $< 1\%$ from the non-turbulent case, implying a minimal impact on guiding.

**Conclusions**

In this work, we have performed the first demonstration of enhanced optical beam transport in air waveguides exposed to realistic outdoor environments, while extending the record guiding length to 100 m. This record was achieved by utilizing a short laser pulse with 210 mJ energy and 1.6 ps duration to drive filamentation, resulting in an air waveguide with a core diameter of $\sim$1 cm. We found that wind has the most deleterious effect on guiding efficiency of all atmospheric effects by causing transverse waveguide drift off the optical axis of the injected probe beam. Even with this limitation, for wind speeds at the Los Alamos field site, we still measured signal enhancement of $1.5\times$ in the waveguide core and $\sim 3$ ms waveguide lifetime.

Our results pave the way for filament-driven air waveguides to be used in challenging outdoor environments for remote sensing, optical communications, and directed energy applications. Wind, although detrimental to air waveguide performance, is not an effect that scales with distance, indicating that an outdoor, quasi-steady-state, kilometer-scale waveguide is achievable with commercially available ultrafast kHz lasers.

## Methods

*Laser systems*

The laser which drives the air waveguides is a Ti:sapphire-based Amplitude Arco X 300 system with output energy 300 mJ, minimum pulse duration 35 fs (FWHM), central wavelength 800 nm, and repetition rate 10 Hz, pumped by frequency-doubled 532 nm Nd:YAG flashlamp-pumped lasers. The beam profile at the laser output is 42 mm (1/$e^2$) diameter with $M^2 \sim 1.7$. For outdoor experiments, we optimize the laser's pulse duration by observing the arrangement and number of filament cores reaching the target at 100 m with burn paper and adjusting the output grating compressor spacing accordingly. From this optimization, we find that a pulse duration of 1.6 ps (FWHM), as measured with a single-shot second-harmonic autocorrelator[33], generated the longest waveguide structures. The beam passes through a 16-segment $l = +1$ spiral phase plate to induce a "donut" intensity profile after the compressor then passes through a two-lens telescope with magnification 5.25:1, such that the beam diameter exiting the telescope is 8 mm, and the energy is 210 mJ from losses due to uncoated optics and quickly-diffracting high-order radial modes introduced by the spiral phase plate.

The 520-nm probe beam is a CW, 4.5 mW, ThorLabs PL251 laser diode. It is coupled into the waveguide with a two-lens telescope placed before the down-collimation telescope to adjust the beam size and focusing. This beam then passes through an 800-nm HR coated dielectric mirror to co-propagate with the 800-nm pulse before also passing through the down-collimation 5.25× telescope. The first two-lens telescope spacing is adjusted to produce the desired beam size and focusing to couple into the waveguide, in our case f/1200. This setup is also illustrated in Fig. 1(b).

*Experimental diagnostics*

Details for the experimental diagnostics are included here. The off-axis camera used to image the guided beam is a ThorLabs Zelux165MU camera with a 0.5-ms exposure time, fast enough to capture an "instant" of the waveguide's thermal evolution. An ensemble of neutral density filters is mounted in front of the camera to prevent sunlight from saturating the camera, with the exact configuration adjusted as needed with position and time of day. A camera lens mounted to the front of the camera is also used to image the guided probe on the Teflon beam block. The photodiode (ThorLabs DET100A2, rise time 35 ns, $\sim$1 cm active-area diameter) rejects filament light with a 520-nm narrow bandpass filter, which limits saturation and allows us to see the temporal evolution of the guide. We isolate the waveguide core with an iris in front of the photodiode to block the light outside of the core region-of-interest of $\sim$1 cm. We use the iris aperture to manually adjust the region-of-interest corresponding to the air waveguide's core diameter at each distance. The camera and oscilloscope are triggered off the laser system to ensure precise timing of the measurements.

The scintillometer (Scintec BLS2000 Neo) is set up with the receiver near the laser source, outside of the container unit, and the transmitter 1 km down the laser path, past our 100 m terminus. Based on properties of the measured light, the scintillometer can retrieve atmospheric turbulence strength (refractive index structure constant $C_n^2$), humidity, pressure, and crosswind speed. Occasionally, the scintillometer was interfered with due to human traffic or scattered infrared filament light, causing corrupted data. This is the cause of the omitted data on September 25. A hot-wire anemometer (Omega HHF-SD1) was used to supplement the scintillometer and measure crosswind speed and air temperature for the September 25 dataset. The anemometer was mounted near the container unit, approximately at beam height, with a one-minute sample rate.

*Air waveguide modeling*

Air waveguide performance is modeled using a paraxial "Beam Propagation Method" which is a two-step Fourier method to calculate monochromatic, linear propagation through an arbitrary index of refraction based on the scalar Helmholtz equation[34]. The solver first initializes a Gaussian field $E(x, y, z = 0) = E_0 e^{-(x^2+y^2)/w_0^2}$ then, for each step, propagates forward by $\Delta z$ in Fourier space:

$$E'(x, y, z + \Delta z) = \mathcal{F}^{-1}\left\{\mathcal{F}\left\{E(x, y, z)\right\} e^{i\Delta z \frac{k_x^2+k_y^2}{\sqrt{k^2-k_x^2-k_y^2}+k}}\right\} \quad (1)$$

Where ($k_x$,$k_y$) are the transverse spatial frequencies of the spatial simulation grid, $k = 2\pi/\lambda$ is the wavenumber, and $\mathcal{F}\{\cdot\}$ ($\mathcal{F}^{-1}\{\cdot\}$) represents the (inverse) Fourier transform. This step represents the "diffractive" part of propagation. This is followed by the "refractive" step that imposes the air waveguide onto the beam:

$$E(x, y, z + \Delta z) = E'(x, y, z + \Delta z) e^{-ik\Delta z(n(x,y,z)-n_0)} \quad (2)$$

Where $n_0 \approx 1 + 2.77 \times 10^{-4}$ is the ambient-density index of refraction for $\lambda = 520$ nm light and $n(x, y, z) = (n_0 - 1)N(x, y, z)/N_0 + 1$ is the spatially dependent refractive index that comes from the estimate density profiles $N(x, y, z)/N_0$, with $n(x, y, z) - 1$ shown in Fig. 5(a) for different $z$. Once (2) is evaluated, the simulation steps forward another $\Delta z$ and the process repeats using the resulting field. The signal enhancement $\epsilon = E_g/E_{ug}$ is then calculated by taking the calculated fields $E(x, y, z)$, and calculating the total energy in the waveguide core with the guide ($N/N_0 < 1$ in the waveguide cladding) and without the guide ($N/N_0 = 1$ everywhere).

Supplemental simulations of performance through turbulence were performed by adding an index generated by the modified von Karman power spectral density[35]:

$$\Phi_n(\kappa) = 0.033 C_n^2 \frac{\exp\{-\kappa^2/\kappa_m^2\}}{(\kappa^2 + \kappa_0^2)^{11/6}} \tag{3}$$

Where $\kappa_m = 5.92/l_0$ is the wavenumber of the inner scale length of turbulence $l_0 = 5$ mm[36] and $\kappa_0 = 2\pi/L_0$ is the wavenumber of the outer scale length of turbulence $L_0 = 10$ m[37]. The phase screens are then generated by taking the square root of $\Phi_n(\kappa)$, multiplying by a randomly seeded complex number of modulus one, and taking the real part of the inverse Fourier Transform.

**Acknowledgements**

The authors acknowledge financial support from Defense Nuclear Nonproliferation Research & Development (Project Number LA25-Air waveguides-PD3Ra) and LANL Laboratory-Directed Research and Development (Project Number 20260916ECR).